\documentclass[apjl,twocolumn]{emulateapj}
\newcommand{\gsim}{\raisebox{-0.13cm}{~\shortstack{$>$ \\[-0.07cm] $\sim$}}~}
\newcommand{\ffm}{4.5 \, \rm  \mu m}
\newcommand{\tfm}{24 \, \rm  \mu m}

\shorttitle{PdBI Cold Dust Imaging of Two $H-[4.5]>4$ Galaxies}
\shortauthors{Caputi et al.}

\begin{document}

\title{P\lowercase{d}BI Cold Dust Imaging of Two Extremely Red $H-[4.5]>4$ Galaxies \\
Discovered with SEDS and CANDELS}


\author{K. I. Caputi\altaffilmark{1},
M. J. Micha{\l}owski\altaffilmark{2},
M. Krips\altaffilmark{3},
J. E. Geach\altaffilmark{4},
M. L. N. Ashby\altaffilmark{5},
J.-S. Huang\altaffilmark{5},
G. G. Fazio\altaffilmark{5},
A. M. Koekemoer\altaffilmark{6},
G. Popping\altaffilmark{1},
M. Spaans\altaffilmark{1},
M. Castellano\altaffilmark{7},
J. S. Dunlop\altaffilmark{2},
A. Fontana\altaffilmark{7},
P. Santini\altaffilmark{7}
}


\altaffiltext{1}{Kapteyn Astronomical Institute, University of Groningen, P.O. Box 800, 9700 AV Groningen, The Netherlands. Email: karina@astro.rug.nl}
\altaffiltext{2}{SUPA, Institute for Astronomy, The University of Edinburgh, Royal Observatory, Edinburgh, EH9 3HJ, UK}
\altaffiltext{3}{Institut de Radio Astronomie Millim\'etrique (IRAM), 300 rue de la Piscine, Domaine Universitaire, F-38406 Saint Martin d'H\`eres, France}
\altaffiltext{4}{Centre for Astrophysics Research, Science \& Technology Research Institute, University of Hertfordshire, Hatfield, AL10 9AB, UK}
\altaffiltext{5}{Harvard-Smithsonian Center for Astrophysics, 60 Garden Street, Cambridge, MA 02138, USA}
\altaffiltext{6}{Space Telescope Science Institute, 3700 San Martin Drive, Baltimore, MD 21218, USA}
\altaffiltext{7}{INAF - Osservatorio Astronomico di Roma, Via Frascati 33, I--00040, Monteporzio, Italy}

\begin{abstract}
We report Plateau de Bure Interferometer (PdBI) 1.1~mm continuum imaging towards two extremely red $H-[4.5]>4$~(AB) galaxies at $z>3$, which we have previously discovered making use of {\em Spitzer} SEDS and {\em Hubble Space Telescope} CANDELS ultra-deep images of the Ultra Deep Survey field. One of our objects is detected on the PdBI map with a $4.3 \sigma$ significance, corresponding to $S_\nu(1.1 \, \rm mm)=0.78 \pm 0.18 \, \rm mJy$. By combining this detection with the {\em Spitzer} 8 and $\tfm$ photometry for this source, and  SCUBA2 flux density upper limits, we infer that this galaxy is a composite active galactic nucleus/star-forming system. The infrared (IR)-derived star formation rate is $SFR\approx 200 \pm 100 \, \rm M_\odot / yr$, which implies that this galaxy is a higher-redshift analogue of the ordinary ultra-luminous infrared galaxies more commonly found at $z\sim2-3$. In the field of the other target, we find a tentative $3.1 \sigma$ detection on the PdBI 1.1~mm map, but 3.7~arcsec away of our target position, so it likely corresponds to a different object. In spite of the lower significance, the PdBI detection is supported by a close SCUBA2 $3.3 \sigma$ detection. No counterpart is found on either the deep SEDS or CANDELS maps, so, if real, the PdBI source could be similar in nature to the sub-millimetre source GN10. We conclude that the analysis of ultra-deep near- and mid-IR images offers an efficient, alternative route to discover new sites of powerful star formation activity at high redshifts.

\end{abstract}


\section{Introduction}
\label{sec-intro}

The powerful star formation and nuclear activity that led to the buildup of massive galaxies through cosmic time have been the subject of many studies. Most of these have focused on the cosmic time period elapsed between redshifts $z\sim 1.5$ and 3, when the cosmic star formation rate density had an overall peak \citep{hop06,beh10}, and the massive galaxy number density had a fast increase \citep[e.g.,][]{fon04,cap06a,sar06,kaj10}. At that time, stellar and nuclear activity were mostly obscured by dust, resulting in a high incidence of ultra-luminous infrared galaxies (ULIRGs). Indeed, a substantial fraction of the most massive galaxies were ULIRGs at $z\sim1.5-3$ \citep{dad05,cap06b}.

The study of powerful star formation activity over the first few billion years of cosmic time ($z>3$) has proven to be more challenging, due to galaxy fainter fluxes, and the gradual decline of the cosmic star formation activity at high $z$.  A notable exception to this challenge is offered by the study of bright sub-/millimetre selected galaxies, whose redshift distribution has a significant tail at $z>3$ \citep[e.g.,][]{war11,mic12}. However, the  sensitivity limits of current sub-/millimetre surveys only allows for the study of the most extreme examples of early dust-obscured star formation, while a plausible population of more typical star-forming ULIRGs at $z>3$ is still to be found.

An alternative approach for finding massive, dust-obscured starbursts at high $z$ consists of selecting bright mid-IR galaxies that are characterised by significantly red colours in their spectral energy distributions (SEDs). These red colours are the result of a redshifted  4000~$\rm \AA$ break and/or significant dust extinction.  For example, different works have shown that optically faint, mid-IR bright galaxies are mostly dusty starbursts lying at $z \gsim 2$, and some also host active galactic nuclei (AGN) \citep[e.g.,][]{yan04,hou05,dey08}.  Restricting this selection to those sources in which the significant flux drop occurs at near-IR wavelengths (observed $\lambda \approx 1-2 \, \rm \mu m$) should produce a redshift distribution biased towards even higher redshifts.

Huang et al.~(2011) reported the existence of four galaxies selected with the {\em Spitzer Space Telescope} Infrared Array Camera \citep[IRAC;][]{faz04}, characterised by colours $H-[3.6]>4.5$ (AB).  Their SED fitting suggests that these galaxies lie at $z\sim4-6$. Similarly, Wang et al.~(2012) analysed the SEDs of 76 IRAC galaxies with $K_s - [3.6]>1.6$ (AB), and found that about half of them are massive galaxies at $z\gsim3$.  

Making use of data from the {\em Spitzer} Extended Deep Survey  \citep[SEDS;][]{ash13} and the {\em Hubble Space Telescope (HST)} Cosmic Assembly Near-infrared Deep Extragalactic Legacy Survey \citep[CANDELS;][]{gro11,koe11},  Caputi et al.~(2012; C12 hereafter) independently searched for these kinds of red galaxies in an area that is part of the UKIRT Infrared Deep Sky Survey \citep{law07}  Ultra Deep Survey (UDS) field. C12 analysed the SEDs of 25 IRAC galaxies characterised by colours $H-[4.5]>4$ (AB), and concluded that between $\sim 45$ and $85\%$ of them are massive galaxies at $z>3$. 

Among the $z>3$ galaxies in C12, six have been detected by the Mid Infrared Photometer for {\em Spitzer} \citep[MIPS;][]{rie04} at $\tfm$, which at $z>3$ traces rest-frame wavelengths $\lambda_{\rm rest} < 6 \, \rm \mu m$, and thus indicates the presence of hot dust. For the brightest sources, this is likely due to the presence of an AGN. Understanding whether these galaxies are simultaneously undergoing a major episode of star formation requires us to follow them up at sub-/millimetre wavelengths, at which the cold-dust continuum emission can directly be probed.

In this work we present PdBI 1.1~mm continuum observations towards the two brightest mid-IR galaxies in the $H-[4.5]>4$ sample analysed by C12. These interferometric observations have allowed us to achieve a spatial resolution of $\sim 1.8 \, \rm arcsec$ and sub-mJy  sensitivities at millimetre wavelengths. Throughout this paper, all quoted magnitudes and colours are total and refer to the AB system \citep{oke83}.  We adopt a cosmology with  $\rm H_0=70 \,{\rm km \, s^{-1} Mpc^{-1}}$, $\rm \Omega_M=0.3$ and $\rm \Omega_\Lambda=0.7$. Stellar masses refer to a Salpeter (1955) initial mass function (IMF) over stellar masses $(0.1-100) \, \rm M_\odot$.

\begin{figure*}
\epsscale{1.1}
\plotone{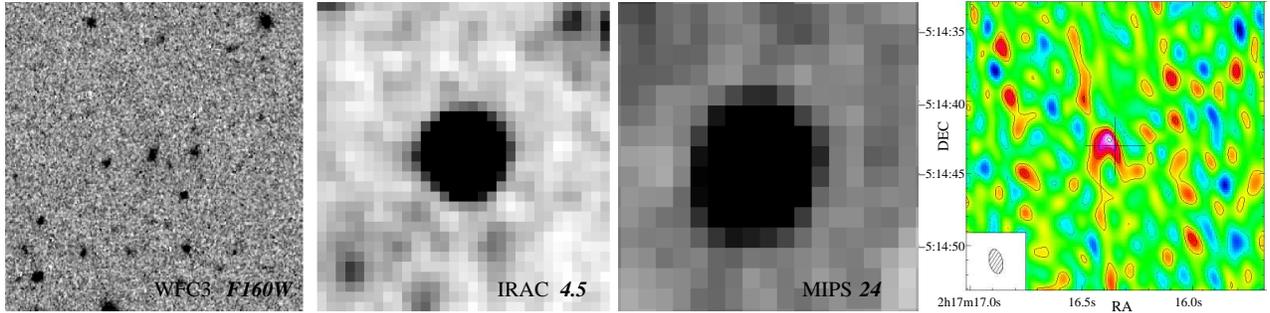}
\caption{Postage stamps of target id \#27564. From left to right: {\em HST} CANDELS f160w;  SEDS IRAC $\ffm$; MIPS $\tfm$;  PdBI clean 1.1~mm  map. The shown field is of $\sim20\times 20$~arcsec$^2$ in all cases. \label{fig_maps27564}}
\end{figure*}

\begin{deluxetable*}{lcccccccc}
\tabletypesize{\scriptsize}
\tablecaption{Photometric properties of our two P\lowercase{d}BI targets. \label{table_targ}}
\tablehead{\colhead{ID} & \colhead{RA (J2000)\tablenotemark{a}} & \colhead{DEC(J2000)\tablenotemark{a}} &  \colhead{F160W} &  \colhead{[4.5]} & \colhead{$S_\nu(\tfm)(\rm \mu Jy)$} & \colhead{$S_\nu(850 \, \rm \mu m)(mJy)$} & \colhead{$S_\nu(1.1 \rm mm)(mJy)$} 
}
\startdata
\#27564 &  02:17:16.35 &  -05:14:43.1 & $24.89\pm0.05$  & $20.39\pm0.10$  & $599 \pm 13$    & $< 2.8 $ & $0.78 \pm 0.18$  \\
\#26857a & 02:17:51.69 & -05:15:07.2 &  $24.39\pm0.14$ & $20.26\pm0.10$  & $334\pm 12$  & $< 2.8 $ & $<1.06$  \\
\#26857b & 02:17:51.62 & -05:15:03.6 & --- & --- & --- &$4.6 \pm 1.4$\tablenotemark{b} & $1.64 \pm 0.53$ 
\enddata
\tablenotetext{a}{The RA and DEC values correspond to the IRAC coordinates, except for \#26857b, for which we quote the PdBI coordinates.}
\tablenotetext{b}{The SCUBA2 $850 \, \rm \mu m$ source centroid is $\sim 3 \pm 4$~arcsec apart from our PdBI source centroid.}
\end{deluxetable*}

\section{Target selection and IRAM P\lowercase{d}BI observations}
\label{sec-data}

Our targets correspond to the two brightest IRAC galaxies reported in C12. Their photometric properties are summarised in Table~\ref{table_targ}. In addition to being bright in all IRAC bands,  these two sources are also bright at $\tfm$, i.e., they have $S_\nu(\tfm)=(599 \pm 13)$ and $(334\pm 12) \, \rm \mu Jy$ for \#27564 and \#26857, respectively. On the other hand, the more recently available SCUBA2 maps have revealed that there is a $3.3 \sigma$ detection with  $S_\nu(850 \, \rm \mu m)=(4.6 \pm 1.4) \, \rm mJy$ within the field of our target \#26857. This field is out of the area covered by SCUBA2 at $450 \, \rm \mu m$. The region around \#27564 has been covered both at $450$ and $850 \, \rm \mu m$, but no $> 2 \sigma$ detection is found within an 8~arcsec distance of our target centroid \citep{gea13}.

The SED fitting analysis based on 17 broad bands ($U$ through $8.0 \, \rm \mu m$) performed by C12 indicates that these two galaxies are at redshifts $z>3$. As for most objects in the C12 sample, the SED fitting solutions are highly degenerate in redshift space, making it very difficult to obtain precise redshift estimates.  However, for these two sources the probability is $P(z>3) \gsim 0.85$, so they can be considered quite secure high-$z$ candidates.

We followed up our two targets with the PdBI in the summer D and C configuration with six antennas, between 24 September and 28 November 2013. We used the WideX correlator tuned to a sky frequency of 265~GHz (corresponding to $\sim1.1 \, \rm mm$), with dual polarization, which produced data over a contiguous 3.6~GHz bandwidth. The weather conditions were reasonable, with water vapour pressures ranging between 1.5 and 3.0~mm. The resulting beam size is of $\sim 1.8$~arcsec, which is comparable to the IRAC $\ffm$ resolution, and the positional accuracy is of around 0.4~arcsec (the PdBI absolute positional accuracy is $< 0.3$~arcsec, but this accuracy is somewhat degraded for faint sources).  The total times on-source were 11.9 and 2.6 hours for targets \#27564 and \#26857, which produced maps with $1\sigma$ depths of 0.18 and 0.53~mJy/beam, respectively. The relative integration times have been decided based on the preliminary SCUBA2 source fluxes/positions available at the time of writing the PdBI proposal.  We performed the data calibration and analysis using the CLIC and MAPPING tasks within the GILDAS software package\footnote{http://www.iram.fr/IRAMFR/GILDAS} \citep{gui00}. The bandpass, complex gain and flux densities have been calibrated with bright ($\gsim 0.5 \, \rm Jy$) standard sources. The main flux calibrator for our observations was  MWC349, which produces a flux accuracy of $\sim 10\%$ at 1.1~mm.

\section{Results}
\label{sec-results}

\subsection{IRAM PdBI maps}
\label{sec-maps}

Figure~\ref{fig_maps27564} shows the clean, full-bandwidth PdBI 1.1~mm map centred at the position of target \#27564, and corresponding CANDELS (f160w) $H$-band, SEDS/IRAC $\ffm$, and MIPS $\tfm$ maps over the same field. The PdBI map shows a robust $4.3 \sigma$ detection centred $0.4$~arcsec apart from the IRAC source centroid, which we can unambiguously identify with our $H-[4.5]>4$ target.

\begin{figure*}
\epsscale{1.1}
\plotone{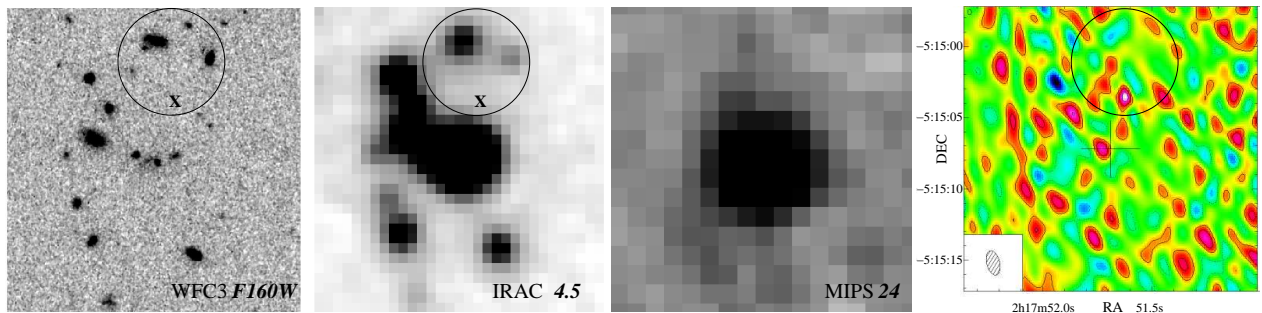}
\caption{Postage stamps of target id \#26857. From left to right: {\em HST} CANDELS f160w;  SEDS IRAC $\ffm$;  MIPS $\tfm$; PdBI clean 1.1~mm  map. The shown field is of $\sim20\times 20$~arcsec$^2$ in all cases. The X-like symbols on the left and middle panels mark the position of the single $>3 \sigma$ detection on the PdBI 1.1~mm map,  which is 3.7~arcsec apart from our target (IRAC) centroid. The circle in each panel is centred at the peak position of the SCUBA2 $\sim3.3 \sigma$ detection, and the radius indicates the positional uncertainty. \label{fig_maps26857}}
\end{figure*}

Figure~\ref{fig_maps26857} shows the corresponding maps for target \#26857. On the PdBI 1.1~mm map a single source appears, with a marginal $3.1\sigma$ detection, located at a distance of 3.7~arcsec from our target centroid. Note that this PdBI source is actually twice as bright as source \#27564 at 1.1~mm, but its detection is less significant due to the considerably shorter integration times.

A $3.1\sigma$ detection on the PdBI map is below the threshold typically considered for robust detections at sub-/millimetre wavelengths \citep[i.e., $\sim3.7\sigma$; e.g.,][]{cop06,wei09}. However, the presence of a  $3.3\sigma$ SCUBA2 source $\sim3 \pm 4$~arcsec away, suggests that the PdBI detection could be the counterpart of the SCUBA2 source (as the positions are consistent within the SCUBA2 positional uncertainty; see fig.~\ref{fig_maps26857}).  Note that the probability associated with a $3.1\sigma$ peak is approximately given by $1-\rm erf(3.1/\sqrt{2}) \approx 0.00195$. About 17 independent PdBI beams are contained within the SCUBA2 positional error circle, which implies that the random probability that a PdBI $3.1\sigma$ peak lies within that area is only of $\sim 0.035$. Assuming a SCUBA2 positional uncertainty radius twice as large as that assumed here would still yield a small probability ($\sim 0.14$). Therefore, these simple statistical arguments suggest that our PdBI $3.1\sigma$ detection is very likely real. However,  we note that these arguments are not totally conclusive, as the probability associated with the S/N ratio given by $1-\rm erf((S/N)/\sqrt{2})$ should only be considered as an approximation.

In any case, the significant separation between the PdBI $3.1\sigma$ source and our target centroid means that it is very unlikely that the millimetre (and sub-millimetre) emission is produced by our $H-[4.5]>4$ IRAC source. We discuss the possibilities for this PdBI detection in detail in Section~\ref{sec-targ26}.

\subsection{Analysis of the Target Multi-wavelength Properties}

\subsubsection{Target \#27564}

\begin{figure*}
\plotone{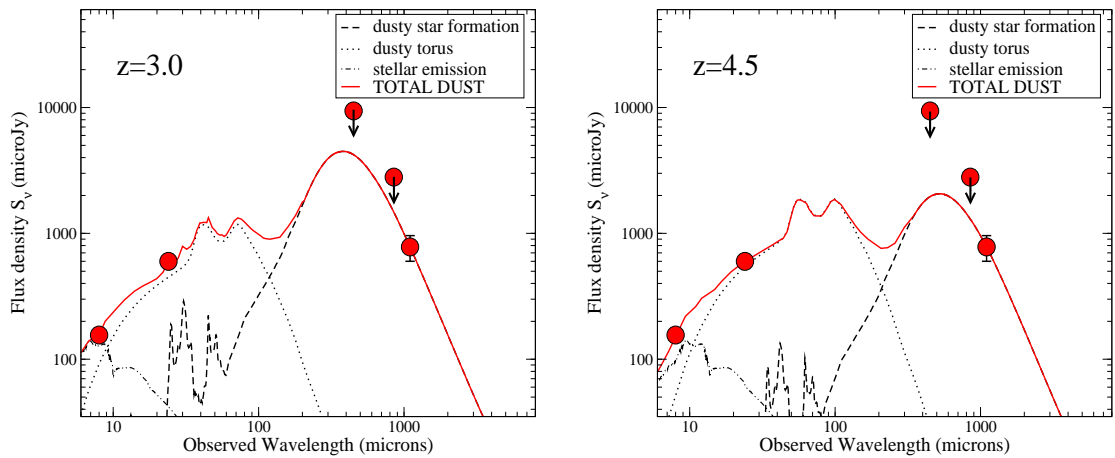}
\caption{Dust IR SED of target id \#27564 at the minimum and maximum most likely redshifts: $z=3$ ({\em left}) and $z=4.5$ ({\em right}). In both panels, the circles correspond to the {\em Spitzer}, SCUBA2 and PdBI photometry. Upper limits correspond to $2 \sigma$ flux densities. These photometric data points are too few to attempt a multi-component dust modelling, so we show an arbitrary dusty AGN torus model, and a pure IR star-forming galaxy model, for an illustrative purpose. Independently of the models chosen, it is evident that an IR star-forming galaxy model alone cannot reproduce simultaneously all the observed IR photometry. The dusty star-forming galaxy model shown here has been taken from the library by Lagache et al.~(2004), while the dusty torus model belongs to the template library by H\"onig \& Kishimoto~(2010).  \label{fig_sedi}}
\end{figure*}

As for target \#27564 the identification with the PdBI detection is unambiguous, we can combine the multi-wavelength information to investigate the dust emission properties of this source.  Figure~\ref{fig_sedi} shows the dust IR SED for this target.  To acknowledge the uncertainties in the redshift determination of this source, we analyse the two extreme values of the redshift interval with maximum probability, i.e. $z=3$ and $z=4.5$. Note that, although there is a non-negligible probability that this source is at higher redshift, we deem that unlikely, as it is detected in the optical $B$ band with $B=26.99 \pm 0.18$~mag (but is not detected in the UDS deep $U$-band images).

With photometry measured in only three bands in the wavelength range $8 \, \rm \mu m - 1.1 \, mm$, and flux density upper limits in other two bands,  we are unable to do a full spectral modelling of our target dust emission. However, from Fig.~\ref{fig_sedi} it is clear that a simple star-forming galaxy model cannot reproduce the sub-/millimetre and mid-IR flux densities altogether. An additional dusty torus component is necessary to reproduce the total IR SED. We have performed an independent, self-consistent SED fitting from the UV through 1.1~mm using the GRASIL code \citep{sil98}, and obtained a similar result: at any of the possible redshifts, no pure star-forming galaxy model can account for the significant excess at mid-IR wavelengths. Therefore, we conclude that our target \#27564 is a composite AGN/star-forming galaxy. 

\begin{figure*}
\plotone{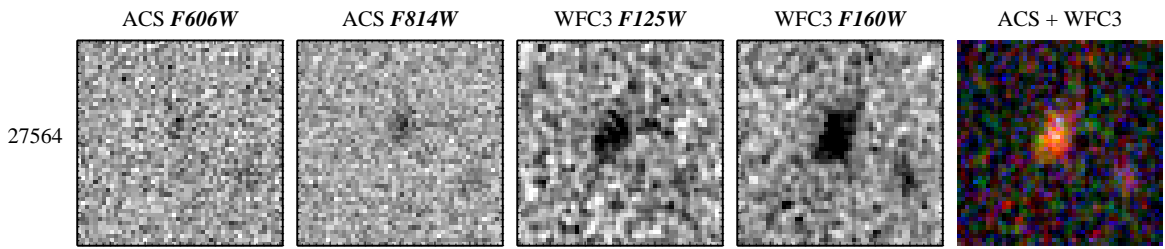}
\caption{Detailed {\em HST} view of target id \#27564 at different wavelengths. The field size in each stamp is of $\sim3\times 3$~arcsec$^2$. \label{fig_hstzoom}}
\end{figure*}

Visual inspection of the {\em HST} images for this source in different bands  (Fig.~\ref{fig_hstzoom}) indicates that this galaxy has a more extended morphology towards longer wavelengths. This suggests the presence of an extended structure with dust-obscured star formation, in correspondence with the millimetre detection. Note that this is not an apparent effect of the lack of sensitivity in the {\em HST} images at short wavelengths. Target \#27564 has magnitudes $H (\rm f160w)=24.89 \pm 0.05$, and $V (\rm f606w)=27.01 \pm 0.23$. The $2\sigma$ depth of the CANDELS/UDS f606w map is $\sim 28$~mag, so the source is well detected in this image, but considerably fainter than in the f160w map.

Based on our PdBI observed flux density at 1.1~mm, we can estimate the total infrared luminosity $L_{\rm IR}^{\rm SFR}$ produced by star formation in our target. The obtained value depends on the adopted IR galaxy template.  Following Micha{\l}owski et al.~(2010b), we scaled different, typical IR galaxy templates to the observed 1.1~mm flux density, obtaining $L_{\rm IR}^{\rm SFR}\sim 0.6-1.7 \times 10^{12}  \, \rm L_\odot$.  For any given template, the derived luminosities $L_{\rm IR}^{\rm SFR}$ are similar at $z=3$ and $z=4.5$, given that the flux dimming at higher $z$ is compensated by the negative $k$ correction.  Considering this range of $L_{\rm IR}^{\rm SFR}$ values, and using the $L_{\rm IR}^{\rm SFR} - SFR$ relation derived by Kennicutt~(1998), we estimate that the obscured SFR of our target is $SFR \approx 200 \pm 100 \, \rm M_\odot/yr$. Note that this SFR would have been largely overestimated if it had been computed starting from the $\tfm$ flux density, which is dominated by the dusty torus emission.

The stellar mass derived for \#27564 is of $\sim 2.5 \times 10^{11} \, \rm M_\odot$ at $z=3$, and $\sim 10^{12} \, \rm M_\odot$ at $z=4.5$, after correcting for the AGN contamination, using a simple power-law component subtraction from the optical through IRAC band photometry \citep[see][]{cap13}. These stellar-mass corrected values are $\sim 30\%$ of the uncorrected ones. Note that, especially the value at $z=4.5$, should still be considered as an upper limit of the real stellar mass. At this redshift, the IRAC bands only trace rest-frame wavelengths $0.6-1.5 \, \rm \mu m$, and the hot-dust power-law component increasingly contaminates the normal galaxy SED up to rest-frame  $\sim 3 \, \rm \mu m$. Thus, observing \#27564  at different wavelengths between observed 8 and $\tfm$ is necessary to really weigh the impact of the AGN power-law component, and derive a fully corrected stellar-mass value.

\subsubsection{Target \#26857}
\label{sec-targ26}

\begin{figure*}
\plotone{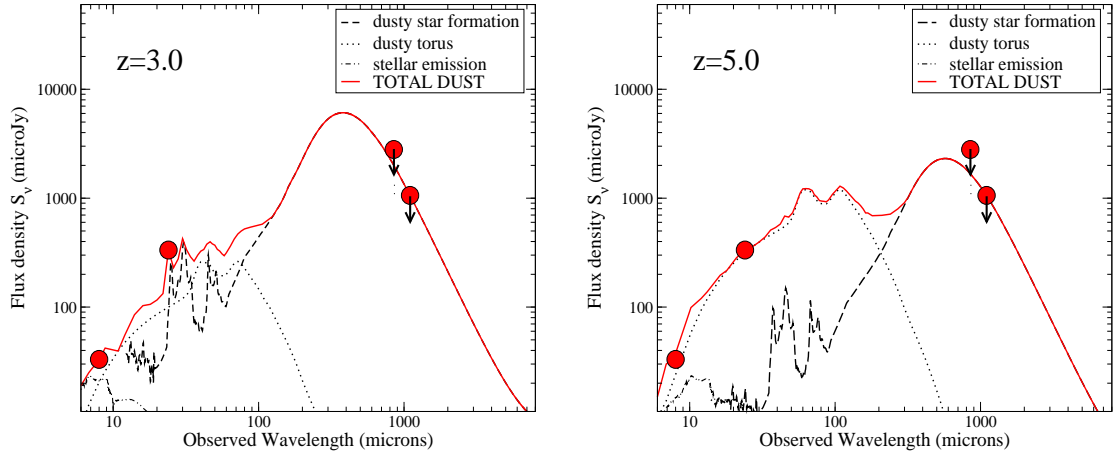}
\caption{Dust IR SED of target id \#26857a at the minimum and maximum most likely redshifts: $z=3$ ({\em left}) and $z=5.0$ ({\em right}). Line styles and symbols are the same as in Fig.~\ref{fig_sedi}.  \label{fig_sedii}}
\end{figure*}

Target \#26857 is not detected on the PdBI map, but we can still attempt to constrain its IR dust SED from the photometric upper limits.  Figure~\ref{fig_sedii} shows this SED at the minimum and maximum most likely redshifts for this source. This source is not detected in any of the UV/optical bands, so in this case we consider $z=5$ as the upper limit for the redshift (note that a higher upper limit will not change the following analysis). At both extreme redshifts, we have normalised the dusty star-forming galaxy template to the PdBI 1.1~mm flux density upper limit, which is the most restrictive one in the far-IR.

From  Fig.~\ref{fig_sedii} it is evident that the pure star-forming galaxy template adjusted to the 1.1~mm photometry could just be sufficient to reproduce the mid-IR ($8$ and $24 \, \rm \mu m$) flux densities, if the galaxy were at $z=3$. Of course, this could have variations depending on the chosen star-forming galaxy template, but in any case a dusty torus component would be of minor importance (unless, of course, the real flux densities at millimetre wavelengths are much lower than the upper limits).  At higher redshifts, a pure star-forming galaxy template is increasingly unable to adjust both the  mid- and far-IR photometry, even for the 1.1~mm the flux density upper limit. Therefore, we can conclude that it is  likely that source \#26857 is also a composite AGN/star-forming system.

As we have discussed in Section \ref{sec-maps}, there is a tentative $3.1\sigma$ detection in the field of this source, but located at 3.7~arcsec from our target. So our target does not appear to be the counterpart of the PdBI/SCUBA2 source. The presence of a  $3.3\sigma$ SCUBA2 $850 \, \rm \mu m$ source, whose  position is consistent with that of our PdBI source within the error bars, suggests that the PdBI detection is likely real (see discussion in \S\ref{sec-maps}). Note that the SCUBA2 and PdBI flux densities are consistent with each other,  considering a typical dust spectral slope of 3.5-4.0.

Interestingly, no counterpart is found at the position of the PdBI detection on the CANDELS $H$-band image, which is remarkable given the depth of the CANDELS UDS maps ($H \approx 27$). Even more surprisingly, no counterpart is found on the deep SEDS $4.5 \, \rm \mu m$ map.  So, if the PdBI source is indeed real, then it will be similar in nature to the presumably rare source GN10 \citep{wan07}, which is very bright at sub-millimetre wavelengths, but extremely faint in the near-IR, and which has been confirmed to be at $z=4.04$ \citep{dan08,dad09}. The PdBI flux density, along with all the photometric upper limits from the $U$ through the $8 \, \rm \mu m$ bands, are consistent with a GN10 SED. We note that these kinds of sources are probably not so rare as initially thought, given that another sub-millimetre source (HDF 850.1) on the same field has been found to have similar properties to those of GN10 \citep{wal12}.

Another possibility could be that the PdBI source is a cold, dusty gas cloud that is associated with our target \#26857. Other cases similar to this have been reported in the literature. For example, Ivison et al.~(2008) have found two sub-millimetre sources associated with a radio galaxy at $z=3.8$, one of which does not have a counterpart in the IRAC bands or shorter wavelengths. They proposed that this sub-millimetre source could be a plume of cold, dusty gas tidally stripped from one of two merging AGN. However,  this plume of cold gas was much closer to its associated AGN ($< 10 \, \rm kpc$) than what our PdBI detection would be from target \#26857 if it were at the same redshift ($\sim 25-30 \, \rm kpc$ at $z\sim 3-4.5$). Therefore, we conclude that the hypothesis that our PdBI detection and target \#26857 are physically associated is much less likely than the possibility that they are different sources.

\section{Constraints on the sub-/millimetre properties of other $H-[4.5]>4$ sources}
\label{sec_othersrcs}

\begin{figure}
\epsscale{1.1}
\plotone{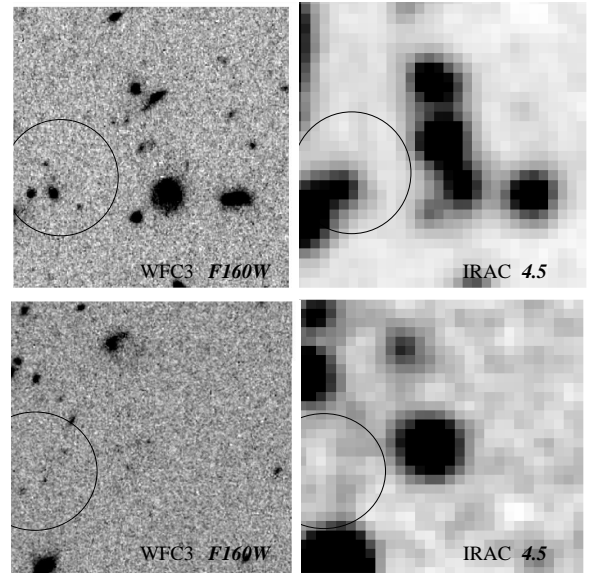}
\caption{Postage stamps of two additional $H-[4.5]>4$ sources. {\em Left}: {\em HST} CANDELS f160w; {\em right:} SEDS IRAC $\ffm$. The shown field is of $\sim20\times 20$~arcsec$^2$ in all cases.  The circle in each panel is centred at the peak position of a SCUBA2 detection, and the radius indicates the positional uncertainty. The SCUBA2 $850 \, \rm \mu m$ detection significances are $\sim4.8 \sigma$ for the source in the top panels, and $\sim3.4 \sigma$ for the source in the bottom panels. The SCUBA2 source in the upper panels is also detected at $450 \, \rm \mu m$ with $3.4\sigma$ significance.  \label{fig_othstamps}}
\end{figure}

\begin{figure*}
\plotone{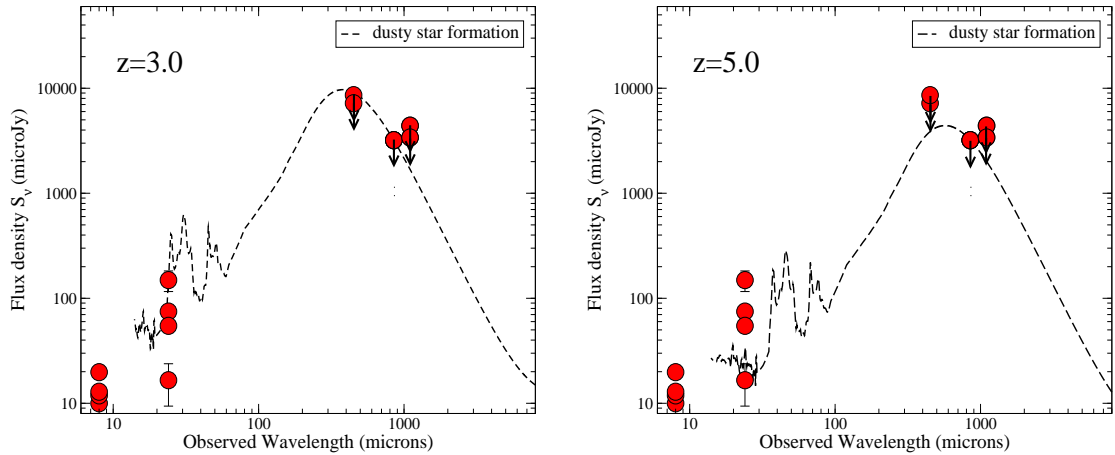}
\caption{Dust IR SEDs of four other $H-[4.5]>4$  sources in the C12 sample, which are $\tfm$ detected. These sources have not been targeted with the PdBI, but we do have SCUBA2 and AzTEC photometric upper limits for them. For clarity, only a pure dusty star-forming galaxy model (dashed line), normalised to the most restrictive of the upper limits, is shown in this case. These plots illustrate that for these other four sources with $H-[4.5]>4$, a pure  dusty star-forming galaxy model is sufficient to account for both the mid-IR photometry and sub-millimetre upper limits at $z=3$. This would not be the case, though, if the real sub-millimetre flux densities were much lower than the upper limits, and/or if the sources were actually at a much higher redshift (right-hand side panel). \label{fig_sedoth}}
\end{figure*}

Four additional $H-[4.5]>4$ sources in the C12 sample are detected at $\tfm$, albeit with fainter fluxes $S_\nu(24 \, \rm \mu m)< 150 \, \rm \mu Jy$. We do not have PdBI observations for them, but we can nevertheless try to constrain their dust IR SEDs, using the SCUBA2 maps, and existing AzTEC 1.1~mm maps for the UDS field \citep{aus10}. 

All of these sources lie in the coverage area of the SCUBA2 $850 \, \rm  \mu m$ maps, and two of them in the $450 \, \rm  \mu m$ coverage region.  For the two sources with only $850 \, \rm  \mu m$  coverage, no $> 3 \sigma$ SCUBA2 detection is found within an 8~arcsec radius. In the fields of the other two sources, with coverage in both SCUBA2 bands, there are respective SCUBA2 $850 \, \rm  \mu m$ ($450 \, \rm  \mu m$)  detections with 4.8 (3.4$\sigma$), and 3.4 ($<2\sigma$) significance.  However, in each case, the SCUBA2 centroid is more than 6~arcsec away from our  $H-[4.5]>4$ source centroid (Fig.~\ref{fig_othstamps}), so it is very unlikely that any of these SCUBA2 sources corresponds to an IRAC $H-[4.5]>4$ source. Note that in the case of the SCUBA2  $850 \, \rm  \mu m$ detection with $4.8\sigma$ confidence (upper panels in Fig.~\ref{fig_othstamps}), there  are two other clear IRAC/WFC3 sources within the sub-millimetre positional uncertainty circle, so one of them (or both) are the likely correct counterparts. In the case of the SCUBA2  $850 \, \rm  \mu m$ detection with $3.4\sigma$ confidence (bottom panels in Fig.~\ref{fig_othstamps}), one would be tempted to associate the SCUBA2 source with the $H-[4.5]>4$ galaxy. However, given the results of our  PdBI observations towards target \#26857, we suspect that this association would likely be wrong.

Therefore, for the study of the IR dust SEDs of all our four additional $H-[4.5]>4$ sources with $\tfm$ detections, we consider that none of them is detected in the SCUBA2 maps, and we use flux density upper limits for the SCUBA2 bands that cover the field of our sources. Figure~\ref{fig_sedoth} shows the dust IR SEDs for these four sources altogether. We only include here a pure dusty star-forming IR galaxy model, normalised to the most restrictive of the sub-millimetre photometric upper limits. These plots illustrate that, for any of these  $H-[4.5]>4$ sources, such a model is sufficient to account for both the mid-IR photometry and sub-/millimetre upper limits at $z=3$.  If the sources were at much higher redshifts (e.g., $z\sim5$), and/or the sub-millimetre fluxes were significantly lower than the SCUBA2 upper limits, then a dusty torus component would be needed. In conclusion, even when we cannot completely exclude the need for a dusty torus, the existing IR photometry indicates that not all of the $H-[4.5]>4$ sources are expected to have an important AGN component, as it is the case for our PdBI target \#27564, and also likely for \#26857.

\section{Discussion}
\label{sec-disc}

Our PdBI detections towards two extremely red $H-[4.5]>4$ galaxies at $z>3$ are important for the following reasons.

$\bullet$ Target  \#27564 has a clear, $4.3 \sigma$-confidence millimetre counterpart, which confirms that this is a massive, AGN/star-forming composite galaxy at high redshifts. The millimetre flux density, which is completely dominated by star formation, indicates that this galaxy has an IR luminosity due to star formation $L_{\rm IR}^{\rm SFR} \approx 0.6-1.7 \times 10^{12} \, \rm L_\odot$, corresponding to $SFR\approx 200 \pm 100 \, \rm M_\odot/yr$. This implies that, from the star-formation point of view, this source is not like the typical hyper-luminous sub-/millimetre sources discovered thus far with single-dish millimetre telescopes at $z\sim 3-4$ \citep[e.g.,][]{mic10}.  Rather, it is a modest ULIRG at $z>3$, such as those more commonly found by IR galaxy surveys at $z\sim 2-3$. Preliminary results on sub-millimetre number counts in deep ALMA observations \citep{kar13,ono14} suggest that, if the redshift distribution of faint sub-millimetre galaxies is comparable to that of the brighter sources currently known, then many more examples of these ordinary ULIRGs should be discovered at $z\sim 3-4$.

$\bullet$ There is a tentative PdBI $3.1 \sigma$ detection at a distance of 3.7~arcsec of our target \#26857. The most likely scenario is that the two sources are unrelated, and the lack of another {\em Spitzer} or {\em HST} counterpart suggests that the PdBI detection corresponds to a new example of a very dusty starburst, like GN10, at high $z$. Our PdBI source also reveals that our $H-[4.5]>4$ galaxy is not the counterpart of the SCUBA2 detection in the same field, as a simple identification of the SCUBA2 source with the brightest IRAC source in the field would suggest.

One could wonder whether the discovery of this new dusty source in the field of our target \#26857 is simply fortuitous. We believe that it likely is not: these red high-$z$ sources tend to be highly clustered \citep[e.g.,][]{tam10,cpk11}, so our finding of a GN10-like candidate source close to our $H-[4.5]>4$ target should perhaps not come as a surprise.  However, this is not necessarily expected for all $H-[4.5]>4$  galaxies, as some of them do not have any close SCUBA2 detection (neither robust nor tentative).

The analysis of the dusty IR SEDs of other $H-[4.5]>4$  galaxies suggests that not all of them are characterised by an important AGN component. At redshifts $z\sim3$, a pure dusty IR star-forming galaxy model is able to reproduce the mid-IR photometry and the sub-/millimetre photometric upper limits in all cases. Therefore, we conclude that our PdBI targets \#27564 and \#26857 could be prototypical of the brightest IRAC $H-[4.5]>4$  galaxies, but not all of them.

As a general conclusion, we argue that the analysis of ultra-deep near and mid-IR data offers an alternative route to discover new sites of powerful star-formation activity over the first few billion years of cosmic time. We also conclude that associations between single-dish sub-millimetre sources and bright IRAC galaxies can be quite uncertain in some cases, and interferometric observations are necessary to study the dust-obscured star formation properties of the $H-[4.5]>4$ galaxies.

\acknowledgments

Based on observations carried out with the IRAM PdBI. IRAM is supported by INSU/CNRS (France), MPG (Germany) and IGN (Spain). Also based on observations made with the {\em Spitzer Space Telescope}, which is operated by the Jet Propulsion Laboratory, California Institute of Technology under a contract with NASA;  the NASA/ESA {\em Hubble Space Telescope}, obtained at the Space Telescope Science Institute; and the James Clerk Maxwell Telescope, operated by the Joint Astronomy Centre on behalf of the UK, Dutch and Canadian scientific councils. The research leading to these results has received funding from the European Commission Seventh Framework Programme (FP/2007-2013) under grant agreement number 283393 (RadioNet3).

We thank Ian Smail for useful discussions on the SCUBA2 maps, and an anonymous referee for a constructive report. MJM acknowledges the support of the UK Science and Technology Facilities Council.

\end{document}